\documentclass [twocolumn, aps, prl, superscriptaddress, showpacs] {revtex4}
\usepackage{graphicx,latexsym}
\begin{document}
\draft
\title {Spin-Polarized to Valley-Polarized Transition in Graphene Bilayers at $\nu=0$ in High Magnetic Fields}
\author {Seyoung Kim}
\affiliation {Microelectronics Research Center, The University of
Texas at Austin, Austin, TX 78758}
\author {Kayoung Lee}
\affiliation {Microelectronics Research Center, The University of
Texas at Austin, Austin, TX 78758}
\author {E. Tutuc}
\affiliation {Microelectronics Research Center, The University of
Texas at Austin, Austin, TX 78758}
\date{\today}
\begin{abstract}
We investigate the transverse electric field ($E$) dependence of the $\nu$=0 quantum Hall state (QHS) in dual-gated
graphene bilayers in high magnetic fields. The longitudinal resistivity ($\rho_{xx}$)
measured at $\nu$=0 shows an insulating behavior which is strongest in the vicinity of $E$=0, and at large $E$-fields.
At a fixed perpendicular magnetic field ($B$), the $\nu$=0 QHS undergoes a transition as a function of $E$, marked by a minimum,
temperature-independent $\rho_{xx}$. This observation is explained by a transition from
a spin polarized  $\nu$=0 QHS at small $E$-fields, to a valley (layer) polarized $\nu$=0 QHS
at large $E$-fields. The $E$-field value at which the transition occurs has a linear dependence on $B$.
\end{abstract}
 \maketitle

Graphene bilayers \cite{novoselov_natphys06} represent an attractive system for electron physics, and
potential device application. This system exhibits a transverse electric field tunable band-gap
\cite{mccannprl06,min}, as evidenced by angle-resolved photoemission \cite{ohta} and transport
measurements \cite{castro07, oostinga}. In a perpendicular magnetic field, graphene bilayers
show quantum Hall states at integer filling factors ($\nu$) multiple of four \cite{novoselov_natphys06, castro07},
owing to spin and valley degeneracy. Electron-electron interaction can lift the Landau level (LL) spin and valley
degeneracy \cite{barlas}, leading to broken symmetry quantum Hall states experimentally observed in single-gated suspended
\cite{feldman09}, and supported \cite{zhao10} graphene bilayers.

We investigate dual-gated graphene bilayers, a device geometry which allows
independent control of the total density and transverse electric field. At a fixed
perpendicular magnetic field ($B$), we observe the emergence of a quantum Hall state
(QHS) at filling factor $\nu=0$ in the presence of a transverse electric field ($E$),
evinced by a large longitudinal resistivity ($\rho_{xx}$) with an insulating behavior,
consistent with the opening of a gap between the electron and hole bands. Interestingly,
as the $B$-field is increased we observe a developing $\nu=0$ QHS at $E$=0,
explained by the Zeeman splitting of the Landau levels at zero energy.
As a function of $E$, the $\nu=0$ QHS undergoes a transition from spin
polarized at small $E$-fields, to valley (layer) polarized at large $E$-fields.

Our samples consist of natural graphite mechanically exfoliated
on a 300 nm SiO$_2$ dielectric layer, thermally grown
on a highly doped $n$-type Si substrate, with an As doping
concentration of $\sim10^{20}$ cm$^{-3}$. Optical inspection
and Raman spectroscopy are used to identify graphene
bilayer flakes for device fabrication. We define metal
contacts using electron beam (e-beam) lithography followed by 50
nm Ni deposition and lift-off [Fig. 1(a)]. A second e-beam
lithography step followed by O$_2$ plasma etching are used to
pattern a Hall bar on the graphene bilayer flake. To deposit the top gate dielectric,
we first deposit a $\sim20\AA$ thin Al layer as a nucleation layer for the atomic layer
deposition (ALD) of Al$_2$O$_3$. The sample is then transferred ex-situ to an ALD chamber.
X-ray photoelectron spectroscopy and electrical measurements confirm the Al layer is
fully oxidized in the presence of residual O$_2$ during evaporation, and the
exposure to ambient O$_2$ \cite{dignam}. Next, a 15 nm-thick
Al$_2$O$_3$ film is deposited using trimethyl aluminum as the Al
source and H$_2$O as oxidizer \cite{kim}, followed by the Ni top gate deposition [Fig.
1(a)]. Longitudinal ($\rho_{xx}$) and Hall ($\rho_{xy}$) resistivity
measurements are performed down to a temperature of $T=0.3$ K,
and using standard low-current, low-frequency lock-in techniques.
Three samples, labeled as A, B, and C, with mobilities of $1500-2400$ cm$^2$/Vs were investigated in this study, all with similar results.

We use Hall measurements to determine the total carrier density ($n_{tot}$) as a function of top
($V_{TG}$) and back ($V_{BG}$) gate voltages, and the corresponding capacitance values.
Equally relevant here is the transverse electric field,
which induces an imbalance between the bottom ($n_B$) and top ($n_T$) layer densities.
Up to an additive constant, $n_{tot}$ and $E$ are related to $V_{TG}$ and
$V_{BG}$ by $n_{tot}=(C_{BG}\cdot V_{BG}+C_{TG} \cdot V_{TG})/e$,
and $E=(C_{BG}\cdot V_{BG}-C_{TG} \cdot V_{TG})/2\varepsilon_0$;
$e$ is the electron charge, $\varepsilon_0$ is the vacuum dielectric permitivity \cite{Efield}.
The $C_{TG}$ values for our samples range between 225 - 270 nF/cm$^2$,
with a dielectric constant $k$=4.2 - 5.

\begin{figure}
\centering
\includegraphics[scale=0.35]{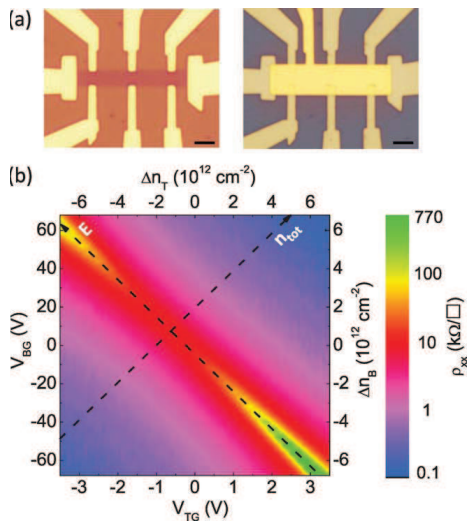}
\caption {\small{(a) Optical micrograph of a dual-gated graphene bilayer, before (left) and after (right) top gate deposition. Both scale bars are 3 $\mu$m.
(b) Contour plot of $\rho_{xx}$ measured as a function of $V_{TG}$ and $V_{BG}$ in sample A.
The right and top axis represent the density change for the back- and top-gates, respectively.}}
\end{figure}

In Fig. 1(b) we show $\rho_{xx}$ measured as a function of $V_{TG}$ and
$V_{BG}$ in sample A, at $T=0.3$ K. The diagonals of constant
$C_{BG} \cdot V_{BG} + C_{TG} \cdot  V_{TG}$ represent the loci of constant
$n_{tot}$ and varying $E$, while diagonals of constant $C_{BG} \cdot V_{BG}-C_{TG} \cdot V_{TG}$
define the loci of constant $E$ at varying $n_{tot}$. The diagonal of $n_{tot}=0$ is
defined by the points of maximum $\rho_{xx}$ measured as a function of $V_{TG}$ at fixed $V_{BG}$
values. In order to determine the $V_{TG}$ and $V_{BG}$ values at which $n_{tot}=0$ and $E=0$,
we consider $\rho_{xx}$ measured along the diagonal $n_{tot}=0$.  The $\rho_{xx}$ shows a
minimum and increases markedly on both sides, thanks to the transverse electric field induced band-gap opening
\cite{mccannprl06,mccann,min,oostinga}. The $\rho_{xx}$ minimum on the $n_{tot}=0$ diagonal defines the $E=0$ point.
Having established a correspondence between ($V_{TG}$, $V_{BG}$) and ($n_{tot}$, $E$),
in the reminder we characterize the bilayers in terms of $n_{tot}$ and $E$.

\begin{figure}
\centering
\includegraphics[scale=0.38]{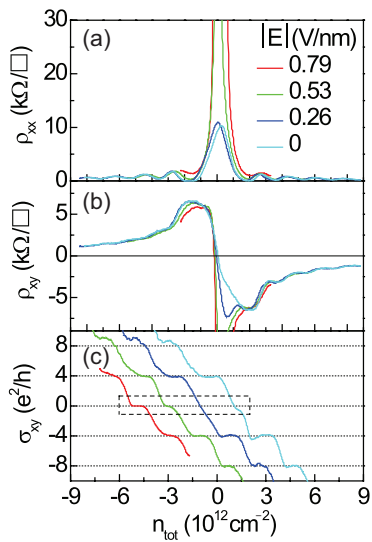}
\caption {\small{(a) $\rho_{xx}$ vs. $n_{tot}$, and (b)
$\rho_{xy}$ vs. $n_{tot}$ measured at $B=18$ T and $T=0.3$ K,
for different $E$-field values in sample A. (c) $\sigma_{xy}$ vs. $n_{tot}$ corresponding to panels (a,b) data,
at different $E$ values, and at $B=18$ T and $T=0.3$ K. The traces are shifted horizontally for clarity.}}
\end{figure}

In Fig. 2(a,b) we show $\rho_{xx}$ vs. $n_{tot}$ and $\rho_{xy}$ vs. $n_{tot}$ respectively,
measured at {\it fixed} $E$-field values, at $B=18$ T and $T=0.3$ K in sample A.
These data are measured by simultaneously sweeping $V_{TG}$ and $V_{BG}$, such that $E$ remains constant.
The data show QHSs, marked by vanishing $\rho_{xx}$ at integer filling factors multiple of four,
consistent with the four-fold degeneracy associated with spin and valley of each Landau level
\cite{novoselov_natphys06, mccannprl06,mccann}. Using the measured $\rho_{xx}$ and
$\rho_{xy}$, we determine the Hall conductivity ($\sigma_{xy}$)
via a tensor inversion, $\sigma_{xy}=\rho_{xy}/(\rho_{xx}^2+\rho_{xy}^2)$.
Figure 2(c) data show $\sigma_{xy}$ vs. $n_{tot}$, measured at $B=18$ T and $T=0.3$ K, and for different values of $E$.
Figure 2(a) data show an increasing $\rho_{xx}$ at $n_{tot}=0$ with
increasing $E$, translating into a Hall conductivity plateau
at $\sigma_{xy}=0$ [Fig. 2(c)], which signals a developing QHS at $\nu=0$ at large $E$.

\begin{figure*}
\centering
\includegraphics[scale=0.6]{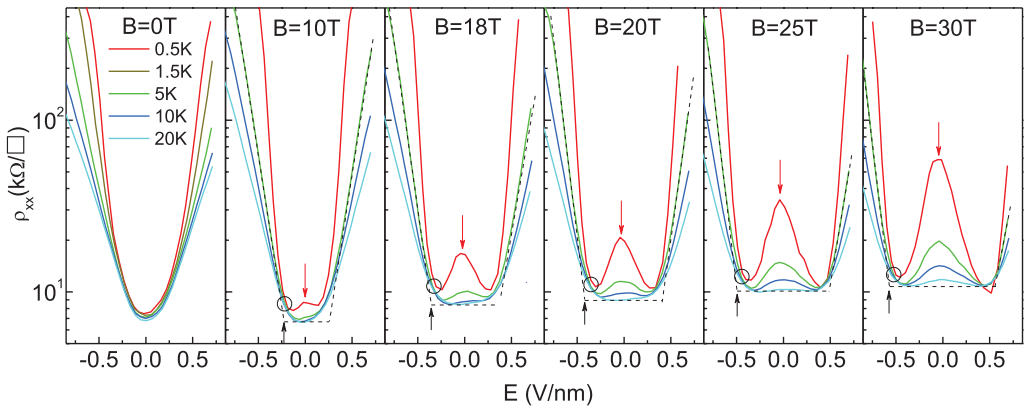}
\caption {\small{$\rho_{xx}$ vs. $E$ measured at $n_{tot}=0$ ($\nu=0$) at different values of the perpendicular $B$-field,
and temperature. At $B=0$, $\rho_{xx}$ shows an exponential dependence on $E$, as well as an insulating behavior at finite $E$,
a consequence of the $E$-field induced band-gap opening in the bilayer. In a perpendicular $B$-field, the onset of the exponential dependence of
$\rho_{xx}$ vs. $E$ (black arrow), which marks the $E$-field induced splitting of the LLs at $\epsilon=0$ increases with the $B$-field.
As the $B$-field increases, $\rho_{xx}$ vs. $T$ shows an insulating state centered at $E=0$, indicating a developing $\nu=0$ QHS
at $E=0$ (red arrow).}}
\end{figure*}

The $\nu=0$ QHS in graphene bilayers at high $E$-fields is explained as follows.
In an applied perpendicular $B$-field the energy spectrum consists of the four-fold, spin and valley degenerate Landau levels (LLs).
At $E=0$ an eight-fold degenerate LL, i.e. the spin and valley degenerate $n=0$ and $n=1$ LLs \cite{mccannprl06,mccann},
exists at energy $\epsilon=0$, the electron-hole symmetry point. The $n=0$, and $n=1$ LL wave-functions are layer polarized \cite{mccannprl06,mccann},
and can be indexed by the layer degree of freedom, in addition to spin. In an applied transverse $E$-field
the eight-fold degenerate LL at $\epsilon=0$ splits into two, four-fold degenerate LLs,
separated the same energy gap ($\Delta$) \cite{mccannprl06,mccann}, which exists between the electron and hole bands at $B=0$.
The higher (lower) energy LLs correspond to the spin degenerate $n=0$ and $n=1$ LLs residing in the layer with higher (lower) on-site energy.

Figure 3 data show $\rho_{xx}$ vs. $E$ measured at different $T$ values, at $n_{tot}=0$. The data is collected
by sweeping $V_{TG}$ and $V_{BG}$ in opposite directions, with sweep rates proportional
to $C_{TG}^{-1}$, and $C_{BG}^{-1}$, respectively. At $B=0$, the $\rho_{xx}$ shows a nearly exponential increase with $E$,
combined with an insulating behavior, a consequence of the $E$-field induced band-gap opening. The $T$-dependence of the $\rho_{xx}$
is weaker than the exponential $\propto e^{\Delta/2k_BT}$ expected for a band insulator, and instead follows more closely a
$\propto e^{(T_0/T)^{1/3}}$ dependence, attributed to variable range hopping between disorder-induced states in the gap \cite{zou,taychatanapat}.
In a perpendicular magnetic field, the $\rho_{xx}$ vs. $E$ data also show an exponential divergence at finite $E$ values, consistent
with the $E$-field induced splitting of the $\epsilon=0$ LLs. However, a closer examination of the $\rho_{xx}$ vs. $E$ data in
high $B$-fields reveals an interesting trend. Let us first consider Fig. 3 data collected at the highest temperature, $T=20$ K.
Unlike the $B=0$ case, the onset of the $\rho_{xx}$ divergence occurs at a {\it finite} $E$-field, which also increases with $B$,
indicating the $E$-field induced LL splitting is suppressed for small transverse $E$-fields. This observation is a direct consequence of the $n=0$
and $n=1$ LLs being layer polarized. Let us assume the transverse $E$-field is applied such that the on-site energy of electrons of the top layer is higher than that on the bottom layer. At filling factor $\nu=0$ the $n=0$ and $n=1$ LLs of the bottom layer will be fully occupied,
while the $n=0$ and $n=1$ LLs of the top layer will be empty. Such LL occupancy will innately place more electrons in the bottom layer,
setting up an internal electric field which opposes the externally applied $E$-field. The magnitude of the internal electric field
is related to the LL degeneracy as
\begin{equation}
E_{int}=4\cdot (e^2B/h)/2\varepsilon_0
\end{equation}

Further examination of Fig. 3 data reveals another interesting finding. In high $B$-fields, $\rho_{xx}$ shows an insulating state centered
at $E=0$, which becomes more pronounced with increasing the $B$-field. This signals a splitting of the $\epsilon=0$ LLs,
and consequently a developing $\nu=0$ QHS at $E=0$, which is attributed to the spin splitting of the $n=0$ and $n=1$ LLs. As the $E$-field
increases $\rho_{xx}$ decreases, and the insulating state weakens. At a fixed $B$-field, the $\rho_{xx}$ vs. $E$ data show a
temperature independent minimum at a critical field $E_c$. For fields higher than $E_c$, the $\rho_{xx}$ shows a diverging dependence on $E$,
a consequence of the $E$-field induced splitting of the $n=0$ and $n=1$ LLs. The $E_c$-field marks a transition at $\nu=0$,
from a spin polarized QHS at small $E$-fields to a valley (layer) polarized QHS at large $E$-fields, in
agreement with several recent theoretical studies which examined the $\nu=0$ phase diagram as a function of transverse
$E$-field, and considering the electron-electron interaction \cite{gorbar10,nandkishore,toke10}. We remark that the
$\rho_{xx}$ vs. $E$ data of Fig. 3 are symmetric for both negative and positive $E$-fields, which indicates the
the disorder is similar for the two layers.

\begin{figure}
\centering
\includegraphics[scale=0.15]{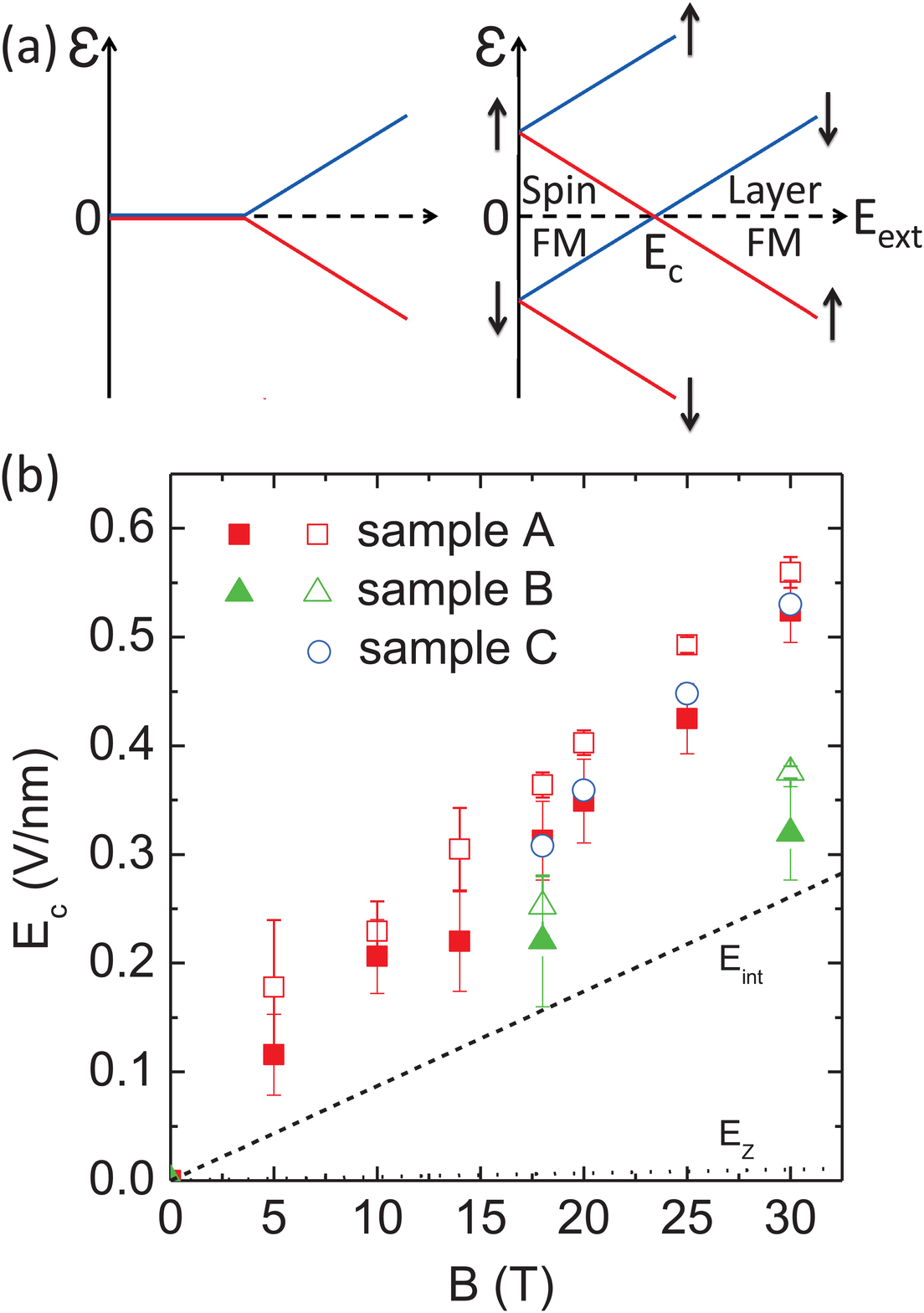}
\caption {\small{(color online) (a) LL energy vs. $E$ dependence neglecting (left panel) and including (right panel)
the electron spin. The red and blue lines denote the LLs corresponding to the bottom and top layer, respectively.
In the absence of spin splitting, the LLs at $\varepsilon=0$ remain degenerate owing to layer polarization (left panel).
When spin (Zeeman) splitting is considered, the $\nu=0$ QHS undergoes a transition at a critical electric field ($E_c$)
from spin polarized (FM) at small $E$ to layer (valley) polarized at large $E$. (b) $E_c$ vs. $B$ measured in three different
samples. The dashed and dotted lines represent $E_{int}$ and $E_Z$ calculated using Eqs. (1) and (2), respectively.}}
\end{figure}

Figure 4(a) shows qualitatively the expected dependence of the $n=0$ and $n=1$ LL energies on the $E$-field. In the absence of
spin splitting  [Fig. 4(a), left panel], the LL layer degree of freedom remains degenerate at finite $E$-field, owing to the
LL layer polarization. In the presence of spin splitting [Fig. 4(a), right panel], the spin down LLs of both layers are occupied,
while the spin up LLs are empty. An applied $E$-field increases (reduces) the energy of the top (bottom) layer LLs, which cross at
a field $E_c$. Figure 4(b) data summarizes the $E_c$ vs $B$ data measured for three samples, marked by different symbols.
We employ two criteria to define $E_c$ using Fig. 3 data. The open symbols in Fig. 4(b) indicate the onset of the $\rho_{xx}$ divergence
at high $E$ fields, shown as a black arrows in Fig. 3. The closed symbols in Fig. 4(b) represent the $E$-fields at which $\rho_{xx}$
is temperature independent, and are marked by circles in Fig. 3. Both criteria yield similar $E_c$ values, with slightly
higher values for the first criterion. It is instructive to compare the the experimental $E_c$ values with two simple calculations.
The first is the electric field ($E_{int}$) required to split the $n=0,1$ LLs when the layer polarization is taken into account [Eq. (1)].
The second is the electric field $E_Z$ at which the electron Zeeman energy ($\Delta_Z$) is equal to the on-site energy difference between the layers:
\begin{equation}
E_Z=g \mu_B B/d
\end{equation}
The $E_Z$ values calculated assuming a $g$-factor of 2, and an inter-layer distance $d=3.4$ {\AA} are represented by the dotted trace in Fig. 4(b); $\mu_B$ is the Bohr magneton. Neglecting interaction, the $\nu=0$ QHS undergoes a transition from spin to valley polarized at an $E$-field equal to $E_Z$.
Examination of Fig. 4 data shows that $E_c$ is much larger than $E_Z$, and comparable albeit larger than $E_{int}$.

We discuss the role of Zeeman splitting on the spin to valley polarized transition. Using $\rho_{xx}$ vs. $E$ at different $B$-fields,
measured at a 48$^{\circ}$ angle between the normal to the sample plane and the magnetic field, we extracted a set of $E_c$
vs. $B$ values similar to Fig. 4(b) data, but with a 1.5 times larger Zeeman splitting. We find that the $E_c$ values remain independent
of the in-plane component of the magnetic field, and are determined only by the $B$-field perpendicular to the sample.
Lastly, we address the role of the edge states. A subtle issue with exfoliated bilayers is the two layers
may not necessarily terminate at the same position, leading to single layer edge states in high magnetic fields.
To test if the edge termination affects Figs. 3 and 4 data, we probed both as exfoliated samples (A),
and samples (B,C) where an O$_2$ plasma etch was used to pattern Hall bars, where both layers
terminate at the same position.

Several theoretical studies have examined the spin to valley polarized transition in graphene bilayers at $\nu=0$.
Gorbar {\it et~al.} \cite{gorbar10} predict a first order phase transition from spin to valley polarized at an $E$-field
of $\simeq$1 mV/nm$\cdot B$[T]. A similar linear $E_c$ vs. $B$ dependence is found in two other studies \cite{nandkishore,toke10},
but with at a larger $E_c$ field, of $\simeq$9 mV/nm$\cdot B$[T]. T\H{o}ke and Fal'ko \cite{toke10} suggest an intermediate, compressible phase
between the spin and valley polarized $\nu=0$ QHSs, with the spin polarized phase collapsing at relatively small electric fields.
Figure 3 data show that the spin polarized phase remains gapped at all fields except for in the vicinity of $E_c$.

A closely similar system to the $\nu=0$ QHS in graphene bilayers, is the $\nu=2$ QHS in double layer GaAs/AlGaAs heterostructures \cite{sawada}.
Depending on the balance between the Zeeman energy, on-site layer energy difference ($\Delta$), and the tunneling energy ($\Delta_t$),
the $\nu=2$ QHS can be either spin or layer polarized, with an intermediate canted spin phase \cite{brey,macdonald}.
The Hartree-Fock theory of the $\nu=2$ QHS \cite{macdonald} shows a first-order transition from spin to layer polarized
when the exchange energy equals the direct (Hartree) energy, a limit reached when $d$ is much smaller than the magnetic length ($l_B=\sqrt{h/eB}$). The $d \ll l_B$ is satisfied up to the highest magnetic fields here, as $d/l_B=0.07$ at $B=30$ T, rendering the $\nu=0$ QHS in graphene A-B bilayers equivalent with the $\nu=2$ QHS in double quantum wells, in the limit of zero tunneling ($\Delta_t=0$), and small Zeeman energy ($\Delta_Z\ll \Delta$). Interestingly, the $d/l_B\simeq0$ limit in GaAs double quantum wells cannot be reached because of limitations associated finite well and barrier widths, finite tunneling, and carrier density.

A recent study of dual-gated, suspended graphene bilayers \cite{weitz} reports a similar transition at $\nu=0$
as a function of transverse electric field, but probed at much lower $E$-fields and up to $B=5.5$ T.
Although the sample mobilities, and the range of $E$-fields and magnetic fields explored in Ref. \cite{weitz} are very different,
remarkably the linear $E_c$ vs. $B$ dependence is in good agreement with the results of this study.

In summary, the $\nu=0$ QHS in dual-gated graphene bilayers in high magnetic field reveals two regimes: at $E=0$,
as a result of the spin splitting, and at large $E$-fields when the system is layer polarized. The $\nu=0$ QHS undergoes a transition from spin to
valley (layer) polarized at a critical electric field ($E_c$), which depends linearly on $B$,
with a slope of 12-18mV/nm$\cdot$T$^{-1}$. Our data, interpreted in the framework of existing theories, suggest the exchange and
direct energies are comparable at $\nu=0$.

We thank A. H. MacDonald, E. V. Castro, D. Tilahun, S. K. Banerjee, I. Jo, and J. H. Seol for many discussions and experimental support.
We acknowledge NRI and NSF (DMR--0819860) for financial support. Part of our work was performed at the National
High Magnetic Field Laboratory, which is supported by NSF (DMR--0654118), the State of Florida, and the DOE.


\begin{thebibliography}{10}
\bibitem{Efield}
To calculate the $E$-field we offset $V_{TG}$ and $V_{BG}$ by the gate biases
required to reach $n_{tot}=0$ and $E=0$.
\end{thebibliography}
\end{document}